\documentclass[twocolumn,showpacs,prl,preprintnumbers,amsmath,amssymb]{revtex4}
\usepackage{bm}
\usepackage{graphicx}% Include figure files
\usepackage{color}
\begin{document}
\title{Delay and distortion of slow light pulses by excitons in ZnO
    }
\author{T.~V.~Shubina$^1$} %\email{shubina@beam.ioffe.ru}
\author{M. M. Glazov$^1$}
\author{N.~A.~Gippius$^{2,3}$}
\author{A.~A.~Toropov$^1$}
\author{D.~Lagarde$^{2,4}$}
\author{P.~Disseix$^2$}
\author{J.~Leymarie$^2$}
\author{B.~Gil$^5$}
\author{G.~Pozina$^6$}
\author{J.~P.~Bergman$^6$}
\author{B.~Monemar$^6$}

\affiliation{$^1$Ioffe Physico-Technical Institute, RAS, St.
Petersburg 194021, Russia} 
\affiliation{$^2$LASMEA, UMR 6602, Universit\'{e} Blaise Pascal, 63177 Aubiere Cedex, France}
\affiliation{$^3$General Physics Institute RAS, Moscow 119991,
Russia} \affiliation{$^4$LPCNO, UMR INSA-CNRS-UPS,
Universit\'{e} de Toulouse, 31077 Toulouse, France}
\affiliation{$^5$GES, UMR5650, Universit\'{e} Montpellier
2-CNRS, Montpellier, France} 
\affiliation{$^6$Department of Physics, Chemistry and Biology, Link\"oping University, S-581 83 Link\"oping, Sweden}

\begin{abstract}

Light pulses propagating through ZnO undergo distortions caused by
both bound and free  excitons.  Numerous lines of bound excitons dissect 
the pulse and induce slowing of light around them, to the extend dependent on their nature. Exciton-polariton resonances determine
the  overall pulse delay and attenuation. The delay time of the
higher-energy edge of a strongly curved light stripe approaches 1.6 ns at
3.374 eV with a 0.3 mm propagation length. Modelling the data of
cw and time-of-flight spectroscopies has enabled us to determine the excitonic parameters, inherent for bulk ZnO.  We reveal the restrictions on these parameters induced by the light attenuation, as well as a discrepancy between the parameters characterizing the surface and internal regions of the crystal.
\end{abstract}


\maketitle

The problem of optical pulse propagation in a medium  has 
attracted attention from the beginning of the last
century \cite{Brillouin}. Loudon \cite{Loudon} studied the
transfer of electromagnetic energy  in a local dielectric with a
resonant absorption line.  A large number of followers
\cite{Bishop,Puri,Branis} extended his analysis to include spatial
dispersion. The experimental time-of-flight study of Chu
and Wong \cite{Chu} of a laser pulse tuned to the bound exciton
(BX) line in GaP   verified the prediction of Garrett and McCumber
\cite{Garrett} on the dramatic variation of group velocity at a
resonance. Their experiments promoted theoretical investigations
of distortion of the temporal  shape of a pulse
\cite{Crisp,Vainshtein,Halevi}. Retardation of exciton-polariton propagation was studied in various semiconductors \cite{Frohlich,Kuwata,Godde}, possessing an isolated resonance, rather than an array of closely situated lines.

Renovation of the interest in slowed light was stimulated by
efforts in quantum information processing \cite{Bigelow}. That
requires a delay significantly exceeding the pulse
duration, while  their ratio is limited by $\sim$4 for quantum
coherent effects \cite{Kasari}. The resonant optical dispersion allows one, in principle, to overcome this limit, however at the price of severe attenuation of the pulse intensity.  
Wide-gap semiconductors, like ZnO and GaN,
promising for various optoelectronics applications, demonstrate strong
excitonic resonances that makes them suitable for  such a light
slowing. Retardation of photoluminescence (PL) by  200
ps was observed near the 3.36-eV BX line in a 1-mm
long ZnO sample \cite{Xiong}. In GaN, the delay was twice higher for the same sample length near BX resonances \cite{Shubina}. In view of
these findings, the question about the distortion of a slowed pulse 
has been revived \cite{Wiersma}. 

\begin{figure} [t]
\includegraphics{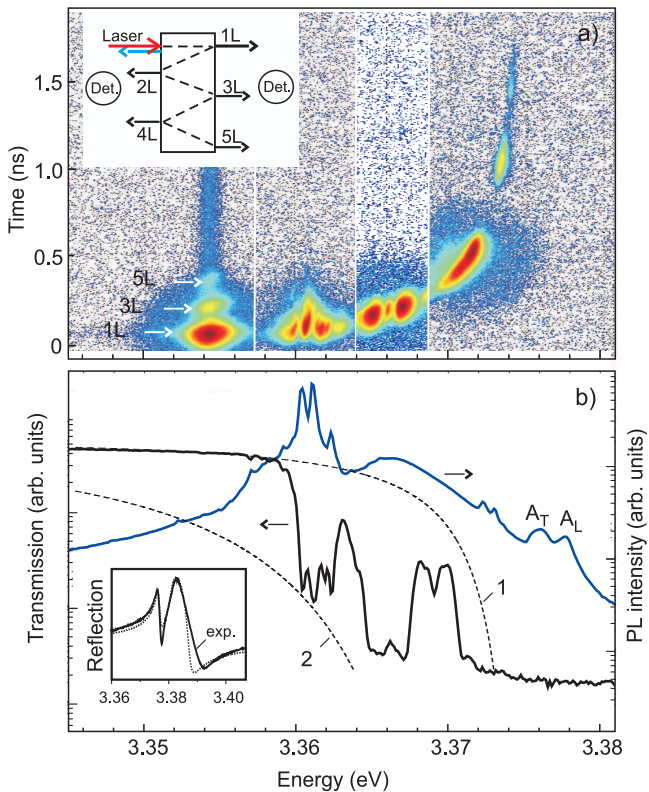}
%  \centering{\epsfbox{Fig1.eps}}
\caption{ \label{f1}(color online) (a) Combined images of four pulses propagating through the 0.3-mm sample (2 K). The boundaries between
them are marked by white lines. The insert presents the scheme of signal detections, where $n\cdot{L}$  denote the series of replicas. 
(b) Spectra of cw PL,  transmission, and reflection (the inset) measured in the same sample. $A_T$ and $A_L$ denote the transverse and longitudinal A exciton emission peaks. Dashed lines show the transmission spectra simulated neglecting the BX lines with $\hbar\Gamma_j$: 1) - 3 $\mu$eV and 2) - 30 $\mu$eV.  
}
\end{figure}

Here, we report on delay and distortion of light pulses by
excitons  in ZnO. The results are novel for two reasons: First, ZnO
possesses numerous lines related to different states of donor bound
excitons. The fine spectrum of these excitons is probed for the first time
by time-of-flight  spectroscopy  via the local distortion of a pulse shape
at different energies. Second, simulation of the general shape of the transmitted pulse permits us to determine the parameters of exciton-polariton resonances, inherent for bulk ZnO. They differ from those given by surface-probing techniques. 
The knowledge of correct parameters is important in many aspects; in particular, they  control the delay and attenuation of a pulse.

The time-of-flight  experiments were performed using the pulses of a 
picosecond tunable laser (Mira-HP, second harmonic).  
A Hamamatsu streak camera with a 2 ps temporal
resolution was exploited to record the time-resolved (TR) images
of the pulses. Their temporal width ($\sim$30 ps) was determined
by instrumental accuracy. Two schemes of the measurements were
exploited, which correspond to the transmission and back-scattering geometries [Fig. 1 (a), the inset]. The measurements were complemented by 
TR and continuous wave (cw)
PL  spectroscopies, as well as cw transmission and reflection
measurements performed at different
temperatures using a tungsten lamp. We investigated high-quality c-plane ZnO samples with the thickness $L$ of 0.3, 0.4, 1, and 2 mm,   supplied by
Tokyo Denpa Co. In the PL spectra, the number (2-4) and intensity of the dominant BX lines varies among the samples, while their energies were almost identical. 

The typical images of the light pulses propagating at different energies are shown in Fig. 1 (a). The leading edge of a curved light stripe, gradually narrowing due to increasing absorption, is observed up to 3.374 eV with the delay  about 1.6 ns, i.e. in the close vicinity of the A exciton. Further shift of the pulse towards higher energies results in its full quenching accompanied by the  increase of BX PL.
In the region of relative transparency, the images
contain series of replicas, arising
due to the light pulse reflection from the crystal boundaries. The
mechanism of their transfer is pure ballistic  \cite{Shubina}. They cover a distance of $n\cdot{L}$, where $n$ is odd (even) number for
the transmission (back-scattering) geometry. The well-defined
temporal intervals between them provide a high accuracy in
determination of the delays.

The images apparently show that the general curvature and delay of
the pulses follow  the optical dispersion controlled by the 
exciton-polariton resonances. The influence of the BX lines is
local, namely: they provide dips cutting the pulses into several
parts and induce extra light retardation nearby. Because even a
very weak line can provide noticeable absorption if a sample is
thick enough, the number of the BX lines resolved by the
time-of-flight  spectroscopy (similar to transmission) is higher than in the PL spectra [Fig. 1 (b)]. The spectral cross-sections of the pulses (Fig. 2) allow
us to reconstruct the fine structure of the BX spectrum which contains at least 18 lines in the 3.356$\div$3.374 eV range (Table I).
In general, these lines can be divided into three groups: i) Two
sets of A and B excitons bound to neutral donors, D$^0$X, separated by a gap of $\sim$4.5 meV. ii) Excitons bound
to ionized donors, D$^+$X, at higher energies. iii) Minor
lines related to the vibrational-rotational $(vr)$ states of the bound
excitons. The lines of the first group produce the strongest
local changes.
 
\begin{figure} [t]
\includegraphics{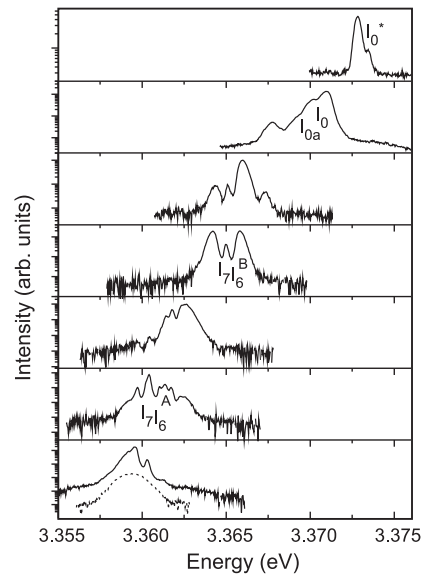}
%  \centering{\epsfbox{Fig2.eps}}
\caption{ \label{f2} Selected spectral cross-sections of the light pulses passing the 0.3-mm sample, made at different delay times from 110 ps (bottom) up to 1200 ps (top). The dips corresponding to several important lines are labelled. The initial light pulse is shown by the dotted curve.
   }
\end{figure}

To model the pulse delay and distortion, the initial  spectrum of the laser pulse $E=E_0$$(\omega)$ in the time moment $t_0$ is approximated as a Gaussian
with the central frequency $\omega_0$. The amplitude and phase of the pulse are given by the Fourier transform at the  boundary of a medium $z=0$ in the linear regime.  In accordance with the
general theory of the electromagnetic field propagation, the
electric field in the spatial point $z=L$ in the moment $t$ can
be described as \cite{Vainshtein}:
\begin{equation}
\label{Efield}
E(L,t)=\int_{-\infty}^\infty E_0(\omega)e^{ik(\omega)L - i\omega t}\frac{d\omega}{2\pi}.
\end{equation}
Here, $k(\omega)=(\omega/c)\sqrt{\varepsilon(\omega)}$ is the complex wave vector of light, $\varepsilon(\omega)$ is the complex dielectric function
of the medium.  

For the medium with several exciton-polariton resonances, 
$\varepsilon(\omega)$ is written as:
$\varepsilon(\omega)=\varepsilon_b+\sum_j X_j$,
where $\varepsilon_b$ is the background dielectric constant and $X_j$ are the contributions of these resonances. 
When the  resonances are assumed to be homogeneous, 
the expression taking into account 
the spatial dispersion can be readily used \cite{Lagois1}. 
With inhomogeneous broadening induced, e.g., by
structural imperfections, each  $X_j$ term can be represented by
the convolution of the line with the Gaussian centered on the same
frequency \cite{Shubina}:
\begin{equation}
\label{response}
X_j=\int\frac{f_j\omega_{0,j}}{\omega_{0,j}+\beta k^2+\xi-i\Gamma_j-\omega}
\frac{1}{\sqrt{\pi}\Delta_j}\exp{\left(\frac{-\xi^2}{{\Delta_j}^2}\right)}d\xi.
\end{equation}
Here, each $j$ resonance is characterized by a frequency $\omega_{0,j}$,
an oscillator strength $f_j$, and a damping term $\Gamma_j$.
$\Delta_{j}$ describes the inhomogeneous width. 
Spatial dispersion is taken into account by the term $\beta
k^2=(\hbar^2/2M_j)k^2$, where the effective  masses $M_j$ are
assumed to be infinite for BX and equal 0.9$m_e$ for each free exciton resonance. 
Equations~\eqref{Efield} and \eqref{response} allow for the spatial dispersion in a simplified way, they are valid for $\omega$ not too close to $\omega_{0,j}$, which is sufficient for our purposes. 
The pulse shapes are calculated from~\eqref{Efield} by means of the Wigner transform.

In our simulations, all noticeable BX lines  were 
taken into account. The $\hbar\Gamma=1$ $\mu$eV and $\hbar\Delta=75$  $\mu$eV characterize these narrow lines. The $f_{BX}$ values corresponding to the observed local distortion are given in Table I.
For the sake of illustration, we present the modelling of the intricate pattern appearing when the central energy of the pulse falls to the gap between the closely situated $I_6$ and $I_7$ lines. The light propagates there as an extremely narrow  $\sim$0.3 meV  stripe  [Fig. 1 (a)]. Figure 3 shows that a twice higher $f$ value taken for the D$^0$X$_B$ transitions ($4\cdot10^{-6}$ vs $2\cdot10^{-6}$) provides an unacceptably long delay and too strong distortion of the pulse.

\begin{table}
\caption{ \label{t1} Parameters of bound excitons in bulk ZnO obtained thought the  pulse distortion analysis. The notation follows Ref. \cite{Meyer}; $X$ denotes unidentified lines.} \begin{ruledtabular}
\begin{tabular}{llllll}
Transition & E (eV) & $f_{BX}$ & Transition & E (eV)  & $f_{BX}$\\
\hline
$I_0^{\ast}(D^+X_A)$  & 3.3734&  $5\cdot10^{-7}$& $X$  & 3.3635 & $5\cdot10^{-7}$\\
$I_{0}(D^+X_{A})$     &  3.3724 & $1\cdot10^{-6}$ &$I_{6}^{vr}(D^0X_{A})$ & 3.3623 & $1\cdot10^{-6}$\\
$I_{0a}(D^+X_{A})$    & 3.3718 & $1\cdot10^{-6}$ & $I_{7}^{vr}(D^0X_{A})$ & 3.3620  & $1\cdot10^{-6}$\\
$X$                   & 3.3705 & $3\cdot10^{-7}$ & $I_{8}^{vr}(D^0X_{A})$& 3.3616 &  $1\cdot10^{-6}$ \\
$X$                   & 3.3695 & $3\cdot10^{-7}$ & $I_{5}(D^0X_{A})$ &  3.3614 & $2\cdot10^{-6}$ \\
$X$                   & 3.3685 & $1\cdot10^{-6}$ &  $I_{6}(D^0X_{A})$ &  3.3608 & $5\cdot10^{-6}$ \\
$I_{5}^{B}(D^0X_{B})$ &  3.3669 & $1\cdot10^{-6}$ & $I_{7}(D^0X_{A})$  & 3.3600 &  $5\cdot10^{-6}$ \\
$I_{6}^{B}(D^0X_{B})$ &  3.3653 & $2\cdot10^{-6}$ & $I_{8}(D^0X_{A})$ &  3.3593  & $5\cdot10^{-9}$\\
$I_{7}^{B}(D^0X_{B})$ &   3.3647 & $2\cdot10^{-6}$ & $I_{9}(D^0X_{A})$ &  3.3566  & $1\cdot10^{-9}$ \\
\end{tabular}
\end{ruledtabular}
 \end{table}
 
 \begin{figure} \includegraphics{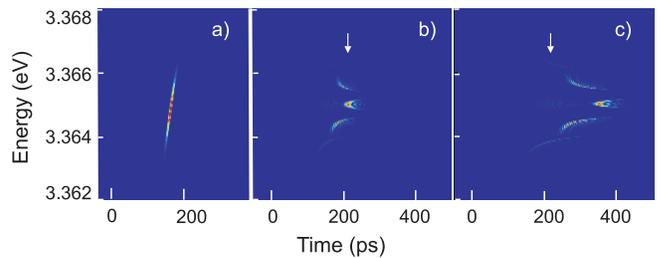}
%  \centering{\epsfbox{Fig4.eps}}
  \caption{ \label{f3} (color online) Simulations of the shape of the 5-ps pulse centered at 3.365 eV   propagating through the 0.3-mm sample: (a) assuming no BX lines;  (b,c) taking into account the BX lines with (b) $f=2\cdot10^{-6}$ and (c) $f=4\cdot10^{-6}$. Arrows mark the experimental delay of the pulse center.
    }
\end{figure}

To determine the parameters of the exciton-polariton
resonances, we consider exclusively the data on
the processes taking place inside the crystal, namely: cw
transmission and pulse propagation. We abstain from using the data of 
reflection and PL spectroscopies, because they probe mostly a region
close to the surface. 
The modelling is simplified owing to the fact that each excitonic parameter is responsible predominantly for a  particular characteristics. For instance, the oscillator strength controls the overall curvature of a pulse via the variation of the group velocity $v_g(\omega)=d\omega/dk$ and, hence,  the different delay $T=L/v_g$ of the pulse constituents. The derived exciton-polariton parameters are collected in Table II (the distant C exciton is neglected).

\begin{table} [b]
\caption{ \label{t2} Exciton-polariton parameters derived from the volume- and surface-probing measurements. 
} 
\begin{ruledtabular}
\begin{tabular}{lcc}
Parameter& A  exciton& B exciton\\
\hline
 &\textit{TR and cw transmission}& \\
$\hbar\omega_{0}$ (eV)& $3.376\pm0.0002$ & $3.382\pm0.001$\\
$f$ & $0.0072\pm0.0002$ & $0.012\pm0.001$ \\
$\omega_{LT}$ (meV)  & $\sim3$ &  $\sim5$ \\
$\hbar\Gamma$ ($\mu$eV)   & $3\pm0.2$  & $3.5\pm0.5$  \\
$\hbar\Delta$ (meV)    &  $<0.5$& $<1$  \\
 &\textit{PL and reflection}& \\
$\hbar\omega_{0}$ (eV)& $3.3758\pm0.0002$ & $3.382\pm0.001$\\ 
$\omega_{LT}$ (meV)  & $1.8\pm0.2$ &  6.5 \\ 
$\hbar\Gamma^{*}$ (meV)  & 0.75 & 1.5 \\

\end{tabular}
\end{ruledtabular}
 \end{table}

The modelling reveals two severe limitations:

1) The damping term $\hbar\Gamma$ must be as small as $\sim$3 $\mu$eV. Higher values would result in the full opaqueness in the transmission spectra in the range where the narrow BX lines are clearly resolved [Fig. 1 (b)]. This is independent on the model used to describe $\varepsilon(\omega)$, because the absorption depends on the imaginary part of the dielectric function, which is determined by this term.

2) The inhomogeneous width $\hbar\Delta$ cannot exceed 0.5 meV for the A exciton resonances. This restriction arises from the observation of the transmitted light at 3.374 eV. 
With stronger broadening, which may occur in crystals of
worse quality or with a temperature rise, the pulse propagating at this energy would  be attenuated up to full disappearance. 
The pulse maximum can be at lower energy due to enhanced absorption at the leading edge.  

\begin{figure} [t]
\includegraphics{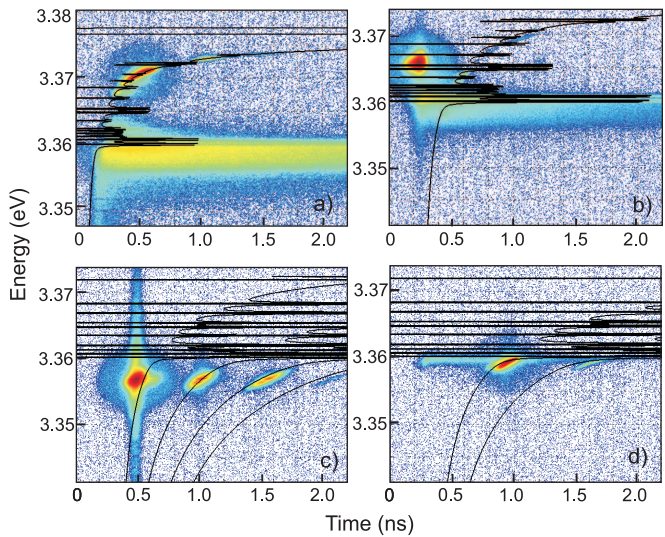}
\caption{ \label{f4} (color online) Light pulses propagating through  the 0.3-mm (a,b)   and  1-mm (c,d) samples. The registration is done at 2 K in the transmission (a,c)
and back-scattering (b,d) geometries. The black lines present the delay
dependencies calculated taking into account the BX transitions.    }
\end{figure}

The critical value of the effective damping parameter,  
yet maintaining the polaritonic modes \cite{Matsushita}, 
is $\hbar\omega_{0}(4f_j\hbar\omega_{0}/Mc^2)^{1/2}\sim1.6$ meV in ZnO for the A exciton, being close to the longitudinal-transverse splitting $\omega_{LT}$ \cite{Klingshirn}.  
Our constraints  based on the signal attenuation are more rigorous. In fact, they do not exclude that the excitonic resonances inside the crystal are homogeneous with $\hbar\Gamma$ of a few $\mu$eV. With these limitations, we have simulated the TR images of the pulses of different energies and the series of replicas, using the same exciton-polariton parameters for all samples (Fig. 4).

Our modelling has revealed obvious contradiction between the results of volume- and surface-probing experiments (Table II). 
The set of the parameters, perfectly fitting the cw and TR
transmission data, results in too sharp peaks in the simulated spectra of reflectivity. 
The successful fitting of the spectrum in Fig. 1 (b), using the model of the homogeneous resonances \cite{Lagois1}, requires the empirical damping term $\hbar\Gamma_{A}^{*}= 0.75$ meV.
Similar result can be obtained assuming the  resonances as inhomogeneous with $\hbar\Delta_A\sim0.9$ meV. However, both variants would provide  too strong  attenuation of the propagating light.
Further, the splitting between the PL peaks, ascribed to transverse and longitudinal A exciton emission, is $1.6-2$ meV, whereas the rough estimation using the bulk parameters gives  $\omega_{LT}\approx3$ meV. There are differences in the resonance energies as well. 
This dependence on the measurement technique is suggestive of certain broadening and deterioration of the excitonic resonances at the surface, which can be induced by modified structural properties and local electric fields \cite{Lagois2,Monemar}. 

In conclusion, our studies exhibit that a strong resonant delay (up to 1.6 ns) of light pulses in ZnO is accompanied by their severe attenuation and shape distortion. Simulation of the  shapes enables us to determine the excitonic parameters, inherent for bulk ZnO, and to establish that the resonances must have the low damping ($\sim3$ $\mu$eV) and limited broadening to allow the light propagation. We believe that this paper will draw attention to the time-of-flight spectroscopy as a new opportunity to investigate the fine structure of BX spectrum. The discrepancy between the surface and volume parameters implies  that only the processes taking place inside the crystal, like the pulse propagation, are suitable to give the true bulk characteristics. 

This work has been supported in part by the RFBR, the Program
of the Presidium of RAS, and the Dynasty Foundation. TVS acknowledges the Universit\'{e} Montpellier 2 for its hospitality.


\begin{references}

\bibitem{Brillouin} For a review, see L. Brillouin, \emph{Wave propagation and group velocity} (Academic, New York, 1960).
\bibitem{Loudon} R. Loudon, J. Phys. A \textbf{3}, 233 (1970).
\bibitem{Bishop} M. A. Bishop and A. A. Maradulin, Phys. Rev. B \textbf{14}, 3384 (1976).
\bibitem{Puri} A. Puri and J. L. Birman, Phys. Rev. Lett. \textbf{47}, 173 (1981).
\bibitem{Branis} S. V. Branis, K. Arya, and J. R. Birman, Phys. Rev. B \textbf{39}, 8371 (1989).
\bibitem{Chu} S. Chu and S. Wong, Phys. Rev. Lett. \textbf{48}, 738 (1982).
\bibitem{Garrett}C.~G.~B. Garrett and D.~E.~McCumber, Phys. Rev. A {\bf 1}, 305 (1970).
\bibitem{Crisp} M. D. Crisp, Phys. Rev. A \textbf{4}, 2104 (1971).
\bibitem{Vainshtein} L. A. Vainshtein, Sov. Phys. Uspekhi \textbf{19 }, 189 (1976).
\bibitem{Halevi} P. Halevi and R. Funchs, Phys. Rev. Lett. \textbf{55}, 338 (1985).
\bibitem{Frohlich} D. Fr\"{o}hlich, A. Kulik, B. Uebbing, A. Mysyrowicz, V. Langer, H. Stolz,
and W. von der Osten, Phys. Rev. Lett. \textbf{67}, 2343 (1991).
\bibitem{Kuwata}M.~Kuwata, T.~Kuga, H.~Akiyama, T.~Hirano, and M.~Matsuoka, Phys.
Rev. Lett. {\bf 61}, 1226 (1988).
\bibitem{Godde} T. Godde, I. A. Akimov, D. R. Yakovlev, H. Mariette, and M. Bayer
Phys. Rev. B \textbf{82}, 115332 (2010).
\bibitem{Bigelow} M. Bigelow, N. Lepeshkin, and R. Boyd, Science {\bf 301}, 200
(2003).
\bibitem{Kasari} A. Kasapi, M. Jain, G. Y. Yin, and S. E. Harris, Phys. Rev. Lett.
\textbf{74}, 2447 (1995).
\bibitem{Xiong}G.~Xiong, J.~Wilkinson, K.~B.~Ucer, and R.~T.~Williams, J.
Phys.:Condens. Matter {\bf 17}, 7287 (2005).
\bibitem{Shubina} T. V. Shubina, M. M. Glazov, A. A. Toropov, N. A. Gippius, A. Vasson,
J. Leymarie, A. Kavokin, A. Usui, J. P. Bergman, G. Pozina, and B. Monemar,
Phys. Rev. Lett. \textbf{100}, 087402 (2008).
\bibitem{Wiersma} D. S. Wiersma,  Nature \textbf{452}, 942 (2008).
\bibitem{Meyer} B. K. Meyer, J. Sann, S. Eisermann, and S. Lautenschlaeger,
Phys. Rev. B \textbf{82}, 115207 (2010).
\bibitem{Lagois1} J. Lagois, Phys. Rev. B \textbf{16}, 1699 (1977).
\bibitem{Matsushita}    M. Matsushita, J. Wicksted, and H. Z. Cummins, Phys. Rev. B \textbf{29}, 3362 (1984).
\bibitem{Klingshirn} C. Klingshirn, J. Fallert, H. Zhou, J. Sartor, C. Thiele,
F. Maier-Flaig, and H. Kalt, Phys. Status Solidi B \textbf{247}, 1424 (2010).
\bibitem{Lagois2} J. Lagois, Phys. Rev. B \textbf{23}, 5511 (1981).
\bibitem{Monemar} B. Monemar, P. P. Paskov, J. P. Bergman, G. Pozina, A. A. Toropov,
T. V. Shubina, T. Malinauskas, and A. Usui,  Phys. Rev. B \textbf{82}, 235202 (2010).


\end{references}
\end{document}